%% file: ms.tex
\documentclass[12pt,preprint]{aastex}

\newcommand{\mdot}{M$_{\odot}$}
\newcommand{\mv}{$M_V$}
\newcommand{\hst}{{\it HST}}
\newcommand{\cha}{{\it Chandra}}

\newcommand{\lx}{$L_{x}$}

\newcommand{\fxfopt}{$F_X/F_{opt}$}
\newcommand{\fopt}{$F_{opt}$}
\newcommand{\ergs}{erg~s$^{-1}$}

\shorttitle{Optical Counterpart to Supersoft Source in M3}
\shortauthors{Edmonds et al.}

\begin{document}

\title{\hst\ Discovery of an Optical Counterpart to the Supersoft X-ray
Source in the Globular Cluster M3\altaffilmark{1}}

\altaffiltext{1}{Based on observations with the NASA/ESA {\it Hubble Space
Telescope} obtained at STScI, which is operated by AURA, Inc. under NASA
contract NAS 5-26555.}

\author{Peter D. Edmonds\altaffilmark{2}, Peter
Kahabka\altaffilmark{3}, \& Craig O. Heinke\altaffilmark{2}}
\altaffiltext{2}{Harvard-Smithsonian Center
for Astrophysics, 60 Garden St, Cambridge, MA 02138;
pedmonds@cfa.harvard.edu, cheinke@cfa.harvard.edu}
\altaffiltext{3}{Sternwarte Universit\"at Bonn, Auf dem H\"ugel 71, D-53121 Bonn,
Germany; pkahabka@astro.uni-bonn.de}

\begin{abstract}

We report the detection, with \hst, of an optical counterpart to the
transient supersoft X-ray source 1E~1339.8+2837, in the globular cluster
M3. The counterpart is found near the faint end of the subgiant branch in
the $V$ vs $V-I$ color magnitude diagram, but is extremely bright in
$U$. Variability is detected over a range of timescales suggesting the
presence of an accretion disk and perhaps also ellipsoidal variations of
the subgiant secondary. The optical colors of the binary are similar to
those of cataclysmic variables recently discovered in 47~Tucanae and
NGC~6397.  We suggest that magnetically channeled accretion may explain the
relatively low X-ray luminosity of this source's supersoft state.

\end{abstract}

\keywords{binaries: general -- globular clusters: individual (M3)
-- techniques: photometric -- X-rays: binaries}

\section{Introduction}
\label{sect.intro}

Supersoft X-ray sources (SSSs) are characterized by high bolometric
luminosities ($10^{36}-10^{38}$ \ergs) and extremely soft X-ray spectra
(blackbody temperatures of 15--80 eV). These characteristics are best
explained by white dwarfs (WDs) that are burning hydrogen-rich matter
accreted onto their surface from a companion star at the rate of about
$10^{-7}$ \mdot\ per year (Kahabka \& van den Heuvel 1997). These accretion
rates are about a hundred times higher than those typically found in
cataclysmic variables (CVs), the more common form of mass accreting binary
containing a WD. Such high accretion rates can be supplied by mass transfer
on a thermal time scale from a slightly evolved companion star (mass
1.3--2.5 \mdot) that is more massive than the WD (Kahabka \& van den Heuvel
1997).

Of the nine known SSSs in the Milky Way (Kahabka \& van den Heuvel 2003),
only one is found in a globular cluster.  A low luminosity X-ray source
(\lx\ $\sim 7 \times 10^{33}$ \ergs) named 1E~1339.8+2837 was discovered
just inside the 33$''$ core radius of M3 (NGC~5272) by {\it Einstein}
(Hertz \& Grindlay 1983; we assume a distance of 10.1 kpc to M3, using
Djorgovski 1993). This source had a much higher luminosity of $2
\times 10^{35}$ \ergs\ when observed by ROSAT in January 1991 and January
1992 (Hertz, Grindlay, \& Bailyn 1993) and a very soft spectrum ($kT
\approx 20$ eV, Hertz et al. 1993; $kT = 36 \pm 20$ eV, Dotani et
al. 1999). This supersoft outburst is 10-100 times less luminous than most
other SSSs (see Kahabka \& van den Heuvel 2003 and Kahabka \& van den Heuvel
1997). The ROSAT luminosity and temperature imply a WD radius of only $9
\times 10^{7}$ cm, suggesting that only part of the WD surface is
undergoing burning. The source has been seen in quiescence (\lx\ $\sim
10^{33}$ \ergs) with a hard X-ray spectrum several times since the
1991/1992 outburst (6/1992, 7/1993, 7/1994, 7/1995 with ROSAT, and 1/1997
with ASCA; Dotani et al. 1999).

Because 1E~1339 is located in a globular cluster its distance and reddening
are well known, helping in modeling of the system. Also, excellent
constraints on the binary parameters are possible if the secondary in the
system is found, combined with good optical photometry. No successful
identification of an optical companion has yet been made, but very few
searches have been reported in the literature. Hertz et al. (1993) report a
star in the ROSAT error circle with UV magnitudes consistent with those of
a horizontal branch star, and Geffert (1998) reports a blue star and a
variable star within a 5$''$ error circle, but provides no further details.

Here, we report the first detection of an optical companion to 1E~1339 in
M3 using the ROSAT X-ray position determined by Verbunt (2001). \hst\
imaging is used to combat crowding near the center of the cluster.  We
briefly discuss the implications of this detection, addressing the origin
of the supersoft X-ray emission, and the parameters of the binary system.
More details about the implications of these results will be given in
Kahabka et al. (2004; in preparation).

\section{Observations, Data Analysis and Results}
\label{sect.obs}

The two archival WFPC2 datasets analyzed for this paper are program GO-5496
obtained in April 1995 (PI: F. Pecci) and program GO-6805 obtained in May
1998 (PI: C. Bailyn).  These observations were obtained when the SSS was
likely in quiescence (see above).  Only a subset of the WF2 images in
GO-5496 were analyzed (those with intermediate length exposures), while all
of the WF2 images for GO-6805 were analyzed (see Table~\ref{tab.data}). In
both of these datasets the candidate for the optical counterpart to 1E~1339
was found on the WF2 chip, and only the images for this chip are analyzed
here.

\subsection{Photometry}
\label{sect.phot}

\begin{figure*}
\epsscale{0.85}
\plotone{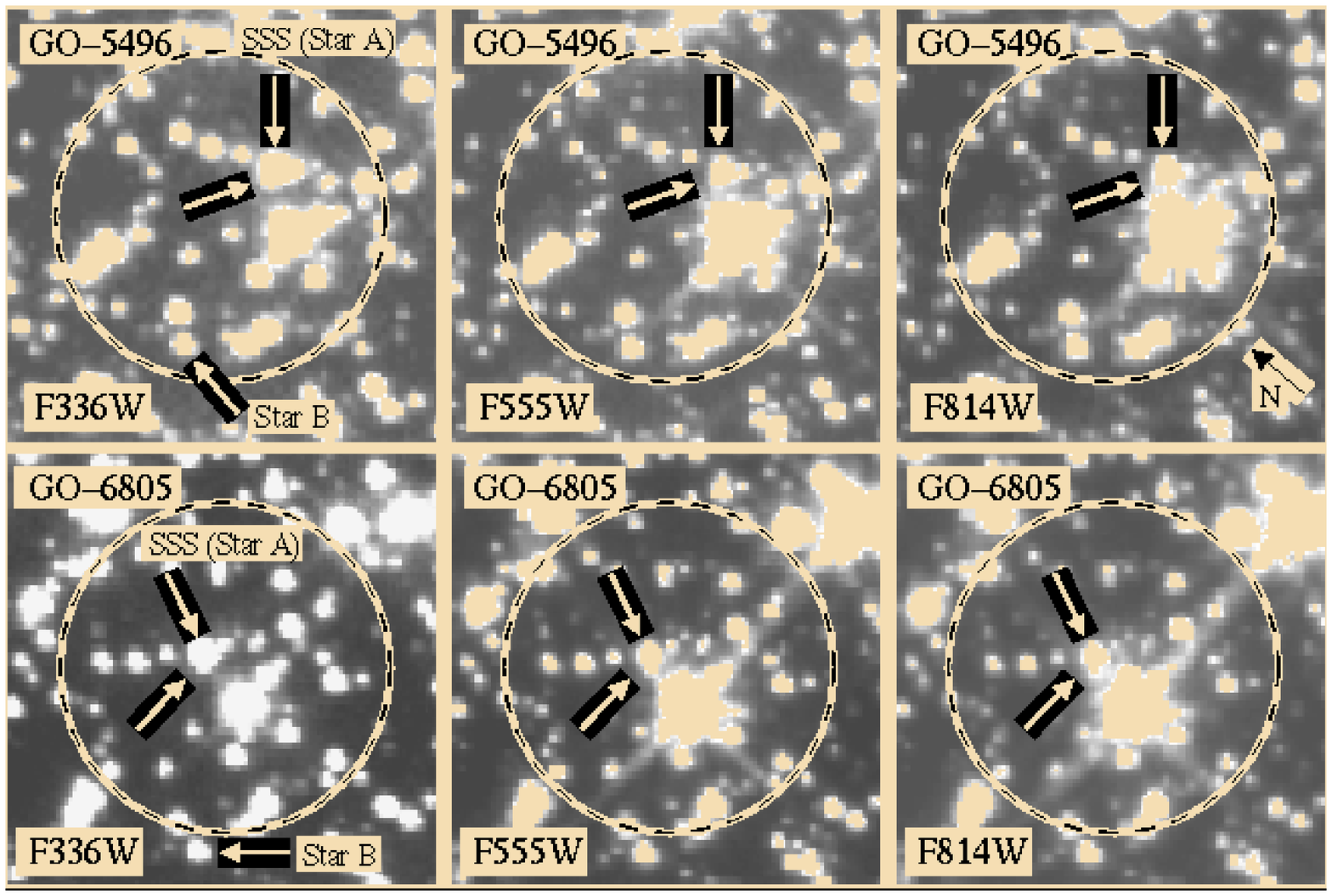}
\caption{Finding charts for the optical counterpart to the SSS (Star
A). F336W ($\sim U$), F555W ($\sim V$) and F814W ($\sim I$) images are
shown for both the GO-5496 (1995) and GO-6805 (1998) datasets. The
dashed-line shows a 2-$\sigma$ (2\farcs2) error circle for the SSS from
Verbunt (2001). Star A, the likely optical counterpart to the SSS is
labeled with two arrows, and Star B, a possible background red giant, is
labeled with a single arrow.}
\label{fig.fchart}
\end{figure*}

For both the GO-5496 and GO-6805 datasets, the F336W, F555W and F814W
images were combined into deep, doubly oversampled images using the drizzle
routines of Fruchter \& Hook (2002). This procedure allowed for the removal
of cosmic rays. Figure~\ref{fig.fchart} shows a small region of the WF2
images in each filter for both datasets. The dashed circle (radius =
2\farcs2, or 2-$\sigma$) is centered on the ROSAT position of 1E~1339
(Verbunt 2001). Note the shift in the position of this circle between the
two \hst\ observing epochs because of $\sim 1''$ uncertainties in the \hst\
astrometry. One star in the X-ray error circle appears from inspection to
be relatively bright in $U$, and is labeled as Star A.  The offsets between
the nominal X-ray position of the SSS and Star A are 0\farcs97
(0.88-$\sigma$) and 0\farcs36 (0.33-$\sigma$) for the GO-5496 and GO-6805
datasets respectively.

To create color magnitude diagrams (CMDs) for the WF2 chip, star lists were
created in each of the three filters and PSF fitting was performed on the
deep images. The instrumental F336W, F555W and F814W magnitudes were then
converted into $U$, $V$ and $I$ magnitudes using the zeropoints and color
corrections of Holtzman et al. (1995). Stars were included in a CMD if they
were detected in either $U$ and $V$ or $V$ and $I$ (the maximum matching
distance between filters was 1.5 oversampled pixels).

\begin{figure}
\epsscale{0.7}
\plotone{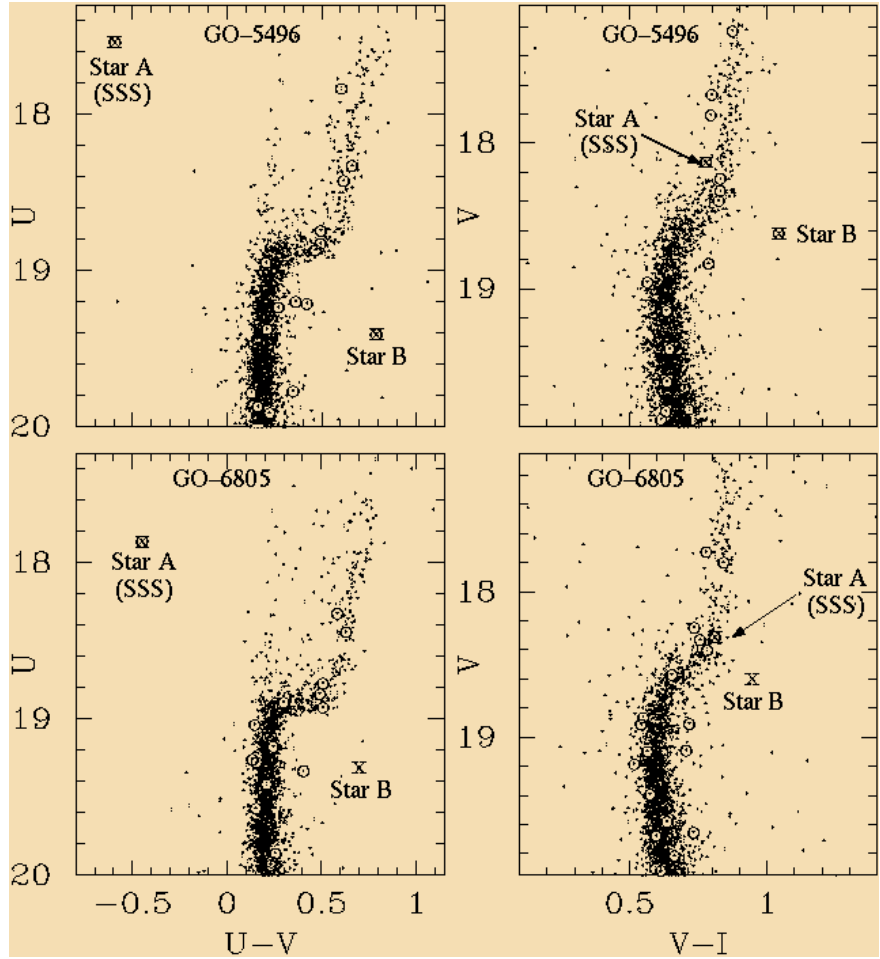}
\caption{$U$ vs $U-V$ and $V$ vs $V-I$ CMDs for the WF2 chip from the
GO-5496 and GO-6805 datasets. Star A, the SSS ID, and Star B, a likely
non-cluster member are labeled with `X's, and stars within 1\farcs5 of the
nominal position of 1E~1339 are circled. Note the different positions of
Star A in the GO-5496 and GO-6805 CMDs.}
\label{fig.cmds}
%\end{figurehere}
\end{figure}

The resulting $U$ vs $U-V$ and $V$ vs $V-I$ CMDs are shown in
Fig.~\ref{fig.cmds}.  Stars within 1\farcs5 of the nominal position of
1E~1339 are circled. The photometric scatter for these objects is slightly
larger than average because the SSS is located in the most crowded region
of the WF2 chip, the area closest to the center of M3. Only two of the
circled objects clearly have unusual colors, and these stars (denoted Stars
A and B) are labeled in each CMD.  Star A is bright and has extremely blue
colors in the $U$ vs $U-V$ CMD, but falls close to the subgiant branch
ridgeline in the $V$ vs $V-I$ CMD (the photometry is given in
Table~\ref{tab.phot}). A comparison of the CMD position of Star A between
the 1995 and 1998 epochs shows evidence for variability, since the star
appears to be slightly fainter and redder in the 1998 dataset in both the
$U$ vs $U-V$ and $V$ vs $V-I$ CMDs (the difference is less obvious with the
redder color). This evidence for variability is confirmed by time series
analysis given in \S~\ref{sect.tseries}.

Star A is not a blue straggler because its $U-V$ colors in both epochs
($-0.6$ and $-0.45$) are much bluer than the mean $U-V$ color of the blue
straggler sequence ($\sim 0.25$), and it lies on the subgiant branch in the
$V$ vs $V-I$ CMD.  This `blue excess' in $U$ and secondary (subgiant)
dominated luminosity in $V$ and $I$ is similar to that found for the CVs in
NGC~6397 (Cool et al. 1998) and 47~Tuc (Edmonds et al. 2003), suggesting
that 1E~1339 is also a CV.  Extra support for this hypothesis is given in
\S~\ref{sect.tseries}.  A key difference between this system and nearly all
of the CVs found in 47~Tuc and NGC~6397 is that Star A includes a subgiant
secondary, with \mv\ ranging between 3.01 and 3.19. The one exception is
AKO~9, the 1.16 day period CV in 47~Tuc (Edmonds et al. 2003).

Star B is located near the outer edge of the 2-$\sigma$ error circle in the
GO-5496 data, and just outside the 2-$\sigma$ error circle in GO-6805. Its
positions in the $U$ vs $U-V$ and $V$ vs $V-I$ CMDs suggest that the star
is a luminous red giant that falls well behind the cluster. Therefore Star
B is unlikely to be responsible for the supersoft X-ray emission.

\subsection{Time Series}
\label{sect.tseries}

\begin{figure}
\epsscale{0.7}
\plotone{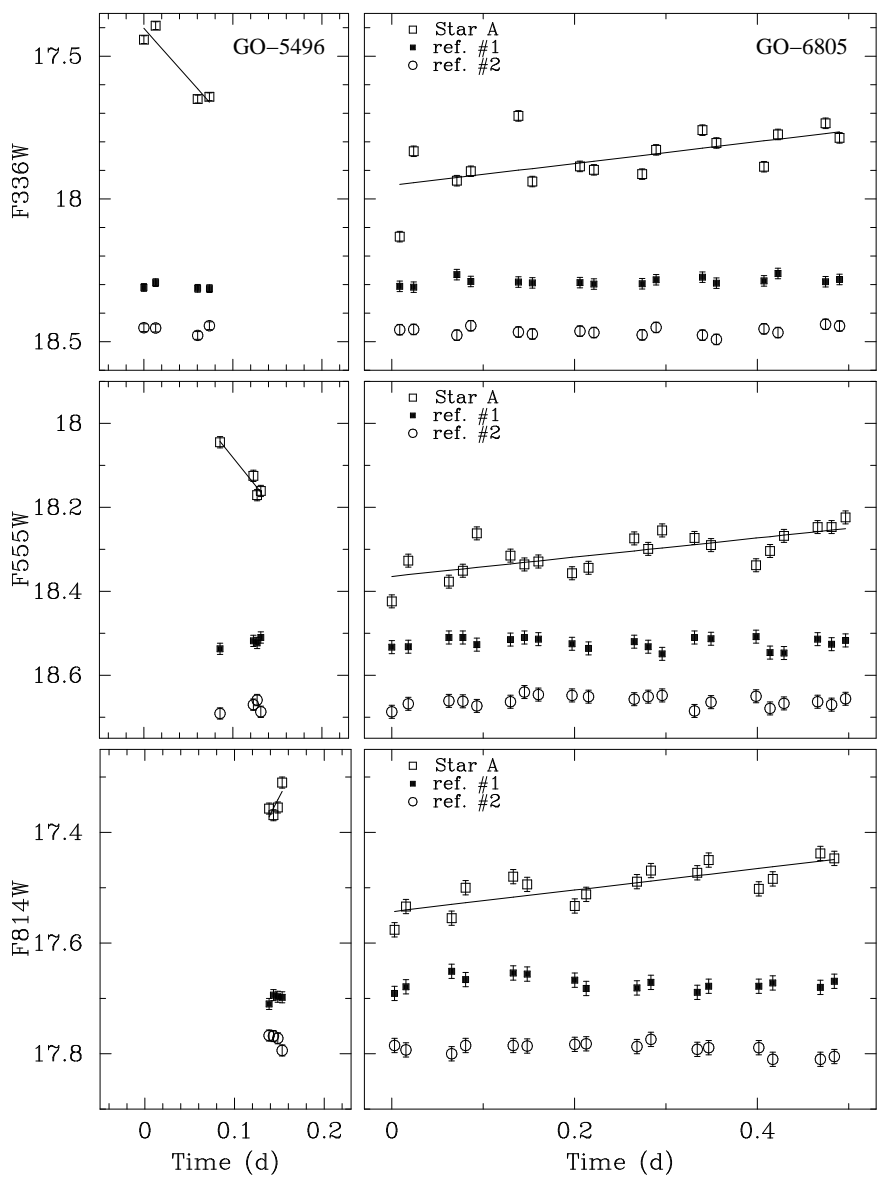}
\caption{Time series for Star A in the GO-5496 and GO-6805 datasets. Time
series for two reference stars with similar magnitudes to those of Star A
are also shown (with appropriate zeropoint shifts to avoid
confusion). Straight-line fits to the time series of Star A are shown, plus
1-$\sigma$ error bars (with approximately the same size as the plotted
symbols in some cases).}
\label{fig.tseries}
%\end{figurehere}
\end{figure}

Figure~\ref{fig.tseries} shows time series in F336W, F555W and F814W for
Star A, with example time series for two stars with similar
magnitudes, offset by $\sim$0.5 mag (for clarity). To derive errors, we
analyzed the time series of 25 stars with magnitudes that are within
several tenths of a magnitude of Star A.  The 1-$\sigma$ error bars
plotted in Fig.~\ref{fig.tseries} show the mean standard deviation for
these 25 stars, with 3-$\sigma$ deviations removed.

Star A is clearly variable.  The standard deviations of the time series for
Star A exceed the mean standard deviations of the 25 reference stars by
factors of 9.8 (F336W), 4.3 (F555W) and 2.6 (F814W) for GO-5496, and by
factors of 5.6 (F336W), 3.2 (F555W) and 3.0 (F814W) for GO-6805. We fit
straight lines ($y=ax+b$) to the data (where $y$ is in units of magnitudes
and $x$ is in units of time measured in days) using least squares fitting
(see Fig.~\ref{fig.tseries}). For the GO-6805 data we derived
$a=-(0.38\pm0.14)$ mag/day (2.8-$\sigma$; F336W), $a=-0.23\pm0.05$ mag/day
(4.7-$\sigma$; F555W), and $a=-0.20\pm0.04$ mag/day (5.1-$\sigma$; F814W).
None of the reference time series in any of the 3 filters had $a > 0.088$
mag. Similar fits are shown for the GO-5496 data, although the short time
series mean that the derived parameters are less useful.

After subtracting the fitted straight lines from the GO-6805 data, we found
that the standard deviation of the F336W time series is still significantly
larger than expected (4.9-$\sigma$), while the excesses for the F555W and
F814W filters are only marginally significant (2.5-$\sigma$ and 2.0-$\sigma$
respectively). Clearly, the GO-6805 variability shows a drift over
the 0.5 day observation added to variations at higher frequencies.

The slow variations shown in Fig.~\ref{fig.tseries} may, in part, be
ellipsoidal variations. These are expected for systems where the secondary
is filling its Roche lobe, and should be detectable, given adequate
signal-to-noise, in systems where the secondary dominates the light in
particular filters (as the SSS does in $V$ and $I$). If the variations are
ellipsoidal in origin, then their approximately linear nature shows we are
observing no more than a quarter of the SSS's orbital period. This implies
that the orbital period of the SSS must be at least $\sim$ 2 days.  Using
the stellar evolution models of Bergbusch \& Vandenberg (1992), Kepler's
third law and the Roche lobe equation of Paczy{\'n}ski (1971) we estimate
that the period of the SSS should be just over 2 days (assuming that the
star's radius does not change significantly from its equilibrium
value). Longer monitoring with \hst\ could easily test for the expected
periodic variations. Ground-based observations will require excellent
seeing or adaptive optics because the SSS is only 0\farcs9 away from a much
brighter neighboring star\footnote{This saturated star is either a red
giant or a horizontal branch star and is identified as the variable star
V224, when using the finding chart of Bakos, Benko, \& Jurcsik
(2000). However, Corwin and Carney (2001) found no evidence for it being
variable. A possible solution to this mystery is that the nearby SSS went
through an outburst and its variability was then so extreme that it
affected measurements of the nearby bright star.}.

Changes in the brightness of the accretion disk may also contribute to the
slow variations.  For example, some contribution from an accretion disk may
be visible in the $V$ vs $V-I$ CMD for GO-5496, where the SSS is bluewards
of the subgiant branch by 0.05--0.1 mag. Also, the magnitude differences
between the GO-5496 and GO-6805 observations are probably caused by some
difference in accretion luminosity.  The shorter term variations are almost
certainly flickering from the accretion disk.

\section{Discussion}
\label{sect.disc}

%
% Absolute mag: 18.13-15.12=3.01; 18.31-15.12=3.19
%

To summarize, the properties of Star A (blue $U-V$ color, variability, and
the close astrometric match with the ROSAT position of the SSS) mean that
it is almost certainly the optical counterpart of 1E~1339.  The optical
colors of the SSS are similar to those of the CVs in 47~Tuc and NGC~6397,
suggesting that 1E~1339 may be a type of CV, consistent with the scenarios
predicted by Hertz et al. (1993) and Dotani et al. (1999).

One popular scenario to allow the high mass transfer rates required for
steady nuclear burning on the surface of the WD is for mass transfer to
occur on a thermal timescale. This is believed to occur in binaries with a
slightly evolved companion star that is more massive than the WD primary
(Kahabka \& van den Heuvel 1997).  With a secondary mass for the SSS of no
more than 0.9 \mdot\ (less depending on how much mass has been lost), it
should be easy to fulfill this requirement by normal stellar evolution in a
primordial binary or an exchange collision between a primordial binary and
a WD.  WDs with masses of $\sim$0.55 \mdot\ are currently being produced in
the cluster (applying the initial-to-final mass relation of Weidemann
2000), and WDs with similar masses will be common.  Therefore, our results
are broadly consistent with expectations of high mass transfer rates to
explain nuclear burning on the surface of the WD primary.  However, a
slightly evolved companion star with a mass between 1.3 and 2.5
\mdot\ is ruled out by these observations, since the only globular cluster
stars in this mass range (besides neutron stars) are blue stragglers.

A useful diagnostic for this system is the \fxfopt\ ratio. We used the
0.5-2.5 keV X-ray flux values of Dotani et al. (1999) from April and June
1995, and January 1997, and \fopt\ $=10^{-0.4V-5.43}$ to derive
\fxfopt=0.9--2.8. These large \fxfopt\ values are consistent with those
found for magnetic (DQ~Her) systems by Verbunt et al. (1997). We speculate
that magnetic behavior might explain why only part of the WD surface
appeared to undergo hydrogen burning in 1991/1992. It has already been
argued that large numbers of magnetic CVs may be found in globular clusters
(Grindlay et al. 1995 \& Edmonds et al. 1999).  We note that studies of
field CVs have shown that some CVs can turn on as SSSs (Greiner et al. 1999
\& Patterson et al. 1998), with luminosities ($10^{35}-10^{36}$ \ergs) that
are lower than those of the original set of Milky Way and Magellanic Cloud
SSSs. One of these, T~Pyx, might be a magnetic CV (Patterson et al. 1998).

\cha\ observations, led by C.O.H., will provide new constraints on the
X-ray spectrum and variability of 1E~1339, and, with a larger sample of
optically identified X-ray sources, should give us excellent (0.1-0.2")
relative astrometry between \hst\ and \cha.  One dataset was obtained in
November 2003 and two will follow in 2004. The second of these \cha\
observations will be coordinated with optical and UV observations with
\hst/ACS in 2004.  Spectroscopic studies of this relatively bright binary
should be possible with \hst/STIS, but will be a challenge for ground-based
telescopes because of the nearby giant star.

\acknowledgments We acknowledge Ron Elsner for useful discussions. 

%\clearpage 

\pagebreak

\input{tab1.tex}

\input{tab2.tex}

\pagebreak

%%--------------------------FIGURE----------------------------------
%%-------------------------------------------------------------------

%%--------------------------FIGURE----------------------------------
%\begin{figurehere}
%%-------------------------------------------------------------------

%%--------------------------FIGURE----------------------------------
%\begin{figurehere}
%%-------------------------------------------------------------------

%---------------------------------------------------------------------------

\end{document}

%% file: tab1.tex
%---TABLE--------------------------------------------------------------------

%\documentclass[preprint]{aastex}
%\begin{document}
\tabletypesize{\footnotesize}

\begin{deluxetable}{lllll}
\tablecolumns{5}
\tablewidth{0pc}
\tablecaption{Summary of optical data}
\tablehead{\colhead{Dataset} &
 \colhead{Observation Date} & \colhead{Camera} & 
 \colhead{Filters} & \colhead{Exposures}}
\startdata

%\cutinhead{Optical results}

GO-5496  & 1995 April 25 & WFPC2 & F336W (`$U$') & 4$\times$800~s \\
         &  & WFPC2 & F555W (`$V$')  & 4$\times$100~s \\
         &  & WFPC2 & F814W (`$I$')  & 4$\times$140~s \\
GO-6805  & 1998 May 14 & WFPC2 & F336W (`$U$') & 16$\times$600~s \\
         &  & WFPC2 & F555W (`$V$')  & 21$\times$60~s \\
         &  & WFPC2 & F814W (`$I$')  & 16$\times$100~s \\

\enddata

\label{tab.data}

\end{deluxetable}

%% file: tab2.tex
%---TABLE--------------------------------------------------------------------

%\documentclass[preprint]{aastex}
%\begin{document}
%\tabletypesize{\scriptsize}
\tabletypesize{\footnotesize}
%\tabletypesize{\tiny}

\begin{deluxetable}{lllccc}
\tablecolumns{6}
\tablewidth{0pc}
\tablecaption{Optical and X-ray data for the companion to 1E~1339}
\tablehead{\colhead{Dataset} &
 \colhead{RA} & \colhead{Dec} & 
 \colhead{$U$} & \colhead{$V$} & \colhead{$I$}  \\
\colhead{} & \colhead{(J2000)} & \colhead{(J2000)} & \colhead{} & 
    \colhead{} & \colhead{} } 
\startdata

%\cutinhead{Optical results}

GO-5496 & 13 42 09.6871\tablenotemark{a} & 28 22 47.046 & 17.53(5) & 18.13(5) &17.35(5)\\
GO-6805 & 13 42 09.7501\tablenotemark{b} & 28 22 47.531 & 17.86(5) & 18.31(5) &17.50(5)\\
ROSAT   & 13 42 09.76\tablenotemark{c}   & 28 22 47.2   & \nodata  & \nodata  & \nodata \\

\enddata

\tablenotetext{a}{Coordinates were derived using metric on \hst\ image u2li0103t}
\tablenotetext{b}{Coordinates were derived using metric on \hst\ image u4r00103r}
\tablenotetext{c}{Coordinates from Verbunt (2001)}

%\tablecomments{bla bla bla}

\label{tab.phot}

\end{deluxetable}